# Understanding the Strain-Dependent Dielectric Behavior of Carbon Black Reinforced Natural Rubber– An interfacial or bulk phenomenon?

Yanhui Huang[*] and Linda S. Schadler

*Department of Material Science and Engineering, Rensselaer Polytechnic Institute, 110 8$^{th}$ street, Troy, New York, USA*

[*]Corresponding Email: huangy12@rpi.edu

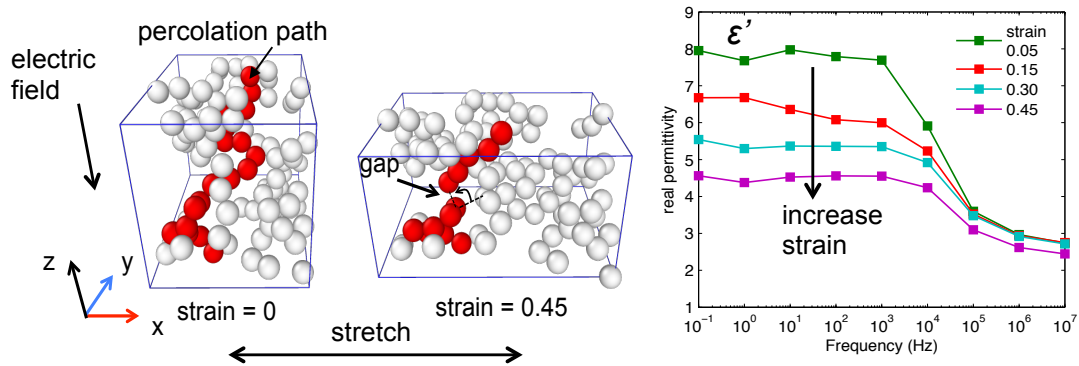

## Abstract

Filler-polymer interactions are one of the keys to understanding the physical properties of polymer composites. These interactions give rise to an interface with specific properties that may have a nontrivial effect on the macroscopic properties of composites. Direct measurement of the interface properties at nanometer scale is usually unavailable. Thus, interface properties are often back calculated from the bulk response using a computational model. However, if the model does not take into account the morphology of the filler dispersion, the results can be misleading. Recently it has been found that the dielectric response of a carbon black filled natural rubber film can change dramatically upon stretching [Huang, *Macromolecules* **49**, 2339 (2016)][1]. In this paper, we will show that this phenomenon can be largely explained by changes in filler cluster connectivity due to strain and is probably not caused by changes in the interfacial interactions. To support the argument, the polarization mechanism of the composites in the measured frequency range is analyzed and numerical models are developed to virtually reproduce the physical phenomenon as a function of strain. As a result, a power-law dependence of dielectric permittivity with strain is derived, which matches closely with the experimental results.

# 1. Introduction

Polymers are often reinforced with nanoparticles to improve the electrical and mechanical properties[2, 3]. For a heterogeneous system, the properties of the filler / matrix interface, and the filler geometry and dispersion morphology are major factors that affect the composite ultimate properties[4-7]. The interface includes a 2D interfacial area and a 3D interfacial region that accounts for the effects extending into the matrix. The properties of the interface depend on the physical and chemical interactions between the two phases and are independent of filler or aggregate geometry and dispersion[5, 8-10]. Filler geometry and dispersion impact properties because they not only determine the volume of the interfacial region, but also influence how mechanical or electrical stress are distributed, at what volume fraction percolation occurs, etc[11, 12]. Directly probing the local interfacial behavior can be difficult due to the small scales and thus, the interface properties are often determined using a computational model that includes the interface and then back calculating the interfacial response from the bulk properties. However, the pitfall of this method is that the change in properties due to the addition of nanofillers is often a combination of interfacial and filler geometry/dispersion/morphology effects. The result can be largely biased if the latter is not accounted for accurately. This is especially true for fillers with fractal geometry or large aspect ratios. To address this issue, finite element based numerical simulations have been developed to include the geometric information derived from the real microscopic images[12-14].

Huang *et al.* recently found that the dielectric response of a carbon black filled natural rubber film can change dramatically upon stretching and it was implied that this is a result of changes in the interfacial properties due to strain[1]. In this paper, we demonstrate that the bulk response can be explained primarily from changes in filler morphology due to strain and thus exclude the changes in interfacial properties as the primary cause of the phenomenon. This paper starts with a brief review of Maxwell-Wagner-Sillars (MWS) polarization, which is believed to dominate the dielectric response of the composite in this case. The impact of filler dispersion on the dielectric permittivity is then discussed. Finite element simulations are used to explicitly model the changes in filler dispersion morphology due to strain and the resulting change in dielectric responses. The change in filler network morphology is analyzed and its effect on the field distribution and dielectric polarization is thoroughly discussed.

## 2. Theoretical Considerations
### 1) Maxwell-Wagner-Sillars (MWS) polarization

Figure 1 shows the measured dielectric spectra of a carbon black filled natural rubber at the volume fraction of 0.13 at different mixing times. The results are taken from Huang [*Macromolecules* **49**, 2339 (2016)], where a N330 carbon black used and the rubber to dicumyl peroxide w/w ratio is 50:1[1]. In all cases, the real permittivity shows a gradual step-increase from $10^7$ Hz to 100 Hz, while the imaginary part exhibits a peak around $10^6$ Hz, and a large increase at 1 kHz and below. The peak in the imaginary permittivity and step-increase of the real part at high frequency is due to the onset of the Maxwell-Wagner-Sillars (MWS) polarization[15, 16]. The sharp rise of imaginary permittivity at low frequency ($f$ < 1 kHz) is caused by dc leakage current, which increases the imaginary permittivity with frequency in a manner of $\varepsilon'' = \sigma_{dc}/\omega$, with $\sigma_{dc}$ being the dc conductivity[17]. But dc conduction has no effect on the real permittivity, as it does not store energy. The electrode polarization can be excluded as no simultaneous increase of real and imaginary permittivity with decreasing frequency are observed[18].

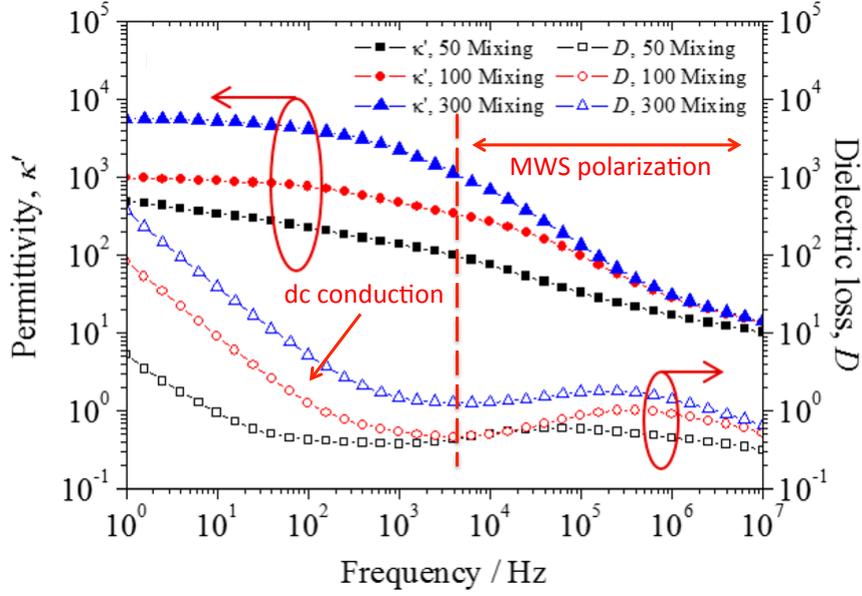

Figure 1 The dielectric spectroscopy of rubber filled with 13 vol% carbon black with different mixing times (i.e. 50, 100, 300) measured at room temperature. Results are taken from Huang [*Macromolecules* **49**, 2339 (2016)]. Figure is adapted with permission. Copyright ACS, 2016.

It is well known that for composites containing conductive fillers in insulating polymers, the dielectric response at radio frequency and below ($f$ < 100 MHz) is largely affected by Maxwell-Wagner-Sillars (MWS) polarization[15, 16]. MWS polarization arises from the accumulation of charges at phase boundaries due to limited conductivity in the polymer matrix. This concentrates the field in the polymer, and thus increases the polarization[17]. At short times, the field partition between two phases is primarily determined by the fast polarization of bound charges, which determines the real permittivity in the high frequency limit, denoted as $\varepsilon_\infty$, and

$$\frac{E_f}{E_m} \sim \frac{\varepsilon_{\infty,m}}{\varepsilon_{\infty,f}}, \quad t < \tau \qquad (1)$$

The subscripts $f$ and $m$ denote the filler and matrix, respectively, and $\tau$ is the time constant of MWS polarization. At long times, free charge (free electron in the case of carbon black) starts to flow and will accumulate at phase boundaries because the conductivities of the two phases are different. The field partition is shifted as a result of charge accumulation until the current density in the two phases equilibrates, and from Ohm's law, this gives

$$\frac{E_f}{E_m} \sim \frac{\sigma_m}{\sigma_f}, \quad t > \tau \tag{2}$$

with $\sigma$ being the conductivity, which states that the long time field partition is primarily determined by the dc conductivity of each phase. Due to abundant free electrons in carbon black,

$$\frac{\sigma_f}{\sigma_m} >> \frac{\varepsilon_{\infty,f}}{\varepsilon_{\infty,m}} \tag{3}$$

which induces giant MWS polarization as the frequency decreases. Figure 2 elucidates this effect by mapping out the field distribution at 10 MHz and 1 kHz respectively for an arbitrary 2D composite containing conductive fillers. It shows that due to the rise of MWS polarization, the field intensification in the polymer matrix increases significantly at the lower frequency.

MWS polarization is a natural result of solving Maxwell's equations in the frequency domain. The polarization has the same form of an ideal Debye dipole relaxation across the spectrum, characterized with a step increase in the real permittivity and a peak in the imaginary part,

$$\varepsilon = \frac{\Delta \varepsilon}{1 + i\omega\tau} + \varepsilon_\infty \tag{4}$$

The relaxation intensity of MWS polarization, $\Delta \varepsilon$, depends on the difference between $\sigma_f / \sigma_m$ and $\varepsilon_{\infty,f} / \varepsilon_{\infty,m}$. The time constant depends on the ratio of the high frequency dielectric constant to the dc conductivity, and for a simple stacking two-layer geometry, the time constant is given by

$$\tau = \frac{\varepsilon_{\infty,1} f_1 + \varepsilon_{\infty,2} f_2}{\sigma_1 f_2 + \sigma_2 f_1} \tag{5}$$

where $\varepsilon, \sigma, f$ are the high frequency dielectric constant, dc conductivity and volume fraction of each phase respectively. The actual composite has more than one time constant due to the complexity of the filler morphology, and the loss peak is usually broader than predicted by (4). Figure 1 shows a loss peak around $10^5 \sim 10^6$ Hz, marking a time constant of $10^{-5} \sim 10^{-6}$ sec, which is expected due to the high conductivity of carbon black. Therefore the MWS polarization in this case starts at a high frequency, with its effect extending into the entire low frequency region, and accounts for the large step-increase in the real permittivity.

Though MWS polarization is also termed "interfacial polarization", it is not a result of interfacial interactions and thus does not explicitly depend on the interfacial area, which distinguishes it from other interfacial effects. This can be most clearly seen from the space charge distribution shown in Figure 2c that the accumulated charges are not uniformly distributed along the interface but mostly concentrated at the upper and lower edges of the connected clusters with opposite signs, forming a large dipole across the entire filler phase. The polarization density is calculated by normalizing the dipole moment to the cluster volume instead of the surface area, which is the same as calculating the polarization density induced by bound charges. This further confirms MWS polarization as a bulk effect.

**2) The dielectric behavior near the percolation threshold**

The polarization in the polymer is critically impacted by the filler morphology as the morphology determines the field distribution and the polarization scales with field. The long-range connectivity of conductive fillers can cause significant field intensification in the polymer matrix, and greatly enhance the polarization and thus the overall permittivity of the composite. (While the field in both polymer and filler are both enhanced, for the convenience of the analysis, we will focus on the effect in the polymer.) As shown in Figure 2b, due to the high conductivity of the filler, the field is most significantly concentrated in the small gap between two connected clusters separated by a distance $d$. Taking the distance between two electrodes as $L$, then the field intensification factor should scale with the ratio of $L/d$, which is related to the size of the interconnected clusters and increases dramatically as the system approaches percolation. Both real and imaginary permittivity of a system near percolation is generally found to increase in a power-law fashion as

$$\varepsilon \propto (f - f_c)^{-s}, f < f_c \tag{6}$$

where $f$ is the volume fraction, $f_c$ is percolation threshold, and $s$ is a factor between 1.1-1.3[11]. From the giant real and imaginary permittivity observed at low frequency in Figure 1, it can be inferred that the field intensification factor is on the order of $10^4$ ~$10^5$, which means that at this point the size of the large interconnected clusters is comparable to the thickness of the sample and the system is close to percolation. The low percolation threshold of ~10 vol% is not surprising for carbon black due to its fractal geometry[19]. This is also supported by the dc conductivity measurement results[1] and is consistent with previous investigations[15, 16].

It is also worth noting that the permittivity and the MWS polarization intensity increases with the mixing times as shown in Figure 1. For the carbon black N330 with moderate structure, the large shearing force involved in the mixing can effectively breakup the dense agglomerates and increase the fractal dimension of the cluster[19]. This actually encourages percolation and the threshold value will be reduced[11], which in turn increases the permittivity as indicated by (6).

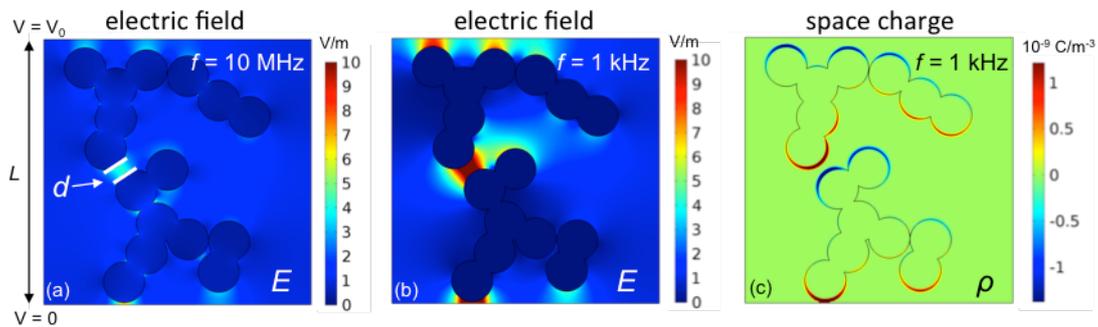

Figure 2 The field distribution of a 2D polymer composite containing conductive fillers that form two interconnected clusters, at (a) 10 MHz and (b) 1 kHz. (c) plots the space charge distribution at 1 kHz. The results are obtained by solving Maxwell's equations at an external field is 1 V/m.

3) **The effect of mechanical strain**

In the original experimental setup, the polymer film is stretched laterally and the voltage is applied along the film plane normal. Upon stretching, the mechanical stress will be concentrated in the filler network due to the higher modulus of the carbon black. The weak van der Waals' interactions between filler clusters may withstand a small elastic deformation, but as strain increases, the contact will break and the load will be transferred to the adjacent rubber matrix, which in turn further increases the separation due to the low modulus of rubber. On the other hand, new contacts may form as the film is under compressive strain in the vertical direction. The net effect requires a more detailed study and here we perform a 3D finite element simulation to explicitly investigate the filler morphology change with strain.

## 3. Finite Element Modeling Methods

1) **Simulating the filler morphology change under strain**

To simplify the problem, we start with an already percolated system and monitor the evolution of filler cluster connectivity with strain. The carbon black is modeled as a cluster that contains 10 interconnected primary spherical particles with a diameter of 50 nm to mimic its fractal structure. The cluster is generated by a diffusion limited aggregation (DLA) method[20], and is denoted as the "primary cluster" to distinguish from higher level aggregations. A larger interconnected cluster consisting of 5 primary clusters is then generated by the DLA method. The boundaries of the simulation box are defined to enclose all particles. The remaining spaces are filled with other primary clusters to maintain a target volume fraction of 10 %. The total number of primary particles in the box is around 80~150. A typical structure is shown in Figure 3.

The input material properties are summarized in Table 1. The connection within the primary cluster is set to be rigid while the inter-cluster connections are modeled explicitly to simulate the detachment process. A cylinder of 10 nm radius is built at each connection point between different primary clusters. The cylinder initially has a large Young's modulus of 4.1 GPa (the value taken from graphite[21]) to account for the van der Waals' interactions between contacted clusters. The modulus then will decrease to 0.1 MPa (a small value compared to polymer) once the average equivalent von Mises stress exceeds 8 MPa (~ 0.2% strain, elastic limit of graphite) to allow large deformation. An additional displacement criterion is imposed by monitoring the distance between two connecting particles. The modulus will change only if the distance increases. Debonding between the filler and polymer is not allowed. The box is stretched along the *x* direction with one boundary fixed and the opposite having a displacement of [*e*X, 0, 0], with *e* being the magnitude of the strain and X being the length in the *x* dimension. The displacement field is solved by a finite element package in COMSOL® based on solid mechanics. Displacement field is calculated, and then the clusters are moved according to their respective displacements to derive the deformed structure.

Because of the large differences in their modulus, the strain is mostly concentrated in the polymer and the strain on the filler is minimal. The calculated strain and stress distribution in the composite are shown in Figure S1 in the supporting information for readers' reference. The tensile stress along the *x* direction was also calculated and plotted in Figure S2 in the supporting information.

The number of contacts per primary cluster is monitored to show the change in connectivity of the structure. Due to the small size of the simulation box, clusters near the boundaries have fewer contacts than the ones in the center, so we only monitor clusters that have at least two contacts in the unstrained condition. As tunneling conduction can occur if two particles are sufficiently close, the contact threshold is set to be 2 nm.

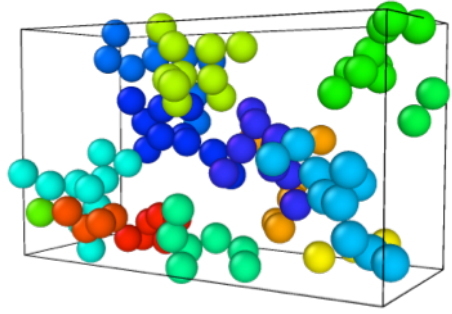

Figure 3 The filler morphology of a generated structure. Each color represents a primary cluster.

Table I The material properties used for the solid mechanics finite element simulation.

| Material | Young's modulus | Poisson's ratio |
| --- | --- | --- |
| carbon black filler | 100 GPa | 0.2 |
| natural rubber | 0.01 GPa | 0.4999 |
| intercluster connection | 4.1/ $10^{-4}$ GPa | 0.2 |

### 2) Simulating the dielectric responses

The dielectric response for each morphology is solved based on solely Maxwell's equations[12] which do not account for any interfacial effects like the electric double layer[5] or interfacial dipoles due to Fermi level equilibration upon contact[8]. To account for the tunneling conduction between fillers at small contact distance, a cylinder of 10 nm radius is built between two particles if their edge to edge distance is smaller than 2 nm. As the tunneling probability decreases exponentially with the distance, the cylinder is assigned a conductivity of

$$\sigma = \sigma_f \exp\left(-\frac{2d}{\gamma}\right) \qquad (7)$$

with $d$ being the tunneling distance, and $\gamma$ being the decay length of the wave function with a value of 2 Å[22]. The input material properties are listed in Table II.

Table II The material properties used for the electrical finite element simulation

| Material | High frequency relative Permittivity | dc Conductivity (S/m) |
|---|---|---|
| carbon black filler[21] | 5 | $10^{-2}$ |
| natural rubber | $3 - 0.01i$ | $10^{-12}$ |
| tunneling connection | $3 - 0.01i$ | $10^{-2} \times \exp(-2d/2 \text{ Å})$ * |

* $d$ is the tunneling distance

## 4. Results and Discussion

Figure 4 shows the number of contacts per primary particle as a function of strain. It can be seen that the net effect is a continuous loss of connectivity as strain increases, but at strain beyond 0.3 the formation of new contact points starts to dominate the process and the number of contacts slightly increases. The field intensification factor, $L/d$, is directly related to the magnitude of the permittivity and can be calculated by finding the shortest path (with minimum crossing distance in the polymer matrix) that bridges the two surfaces along $y$ or $z$ direction. Figure 5 shows an evolution of the filler morphology as a function of strain. The red particles outline the shortest path from the top to bottom surface. A small separation gap is observed at a strain of 0.15 and is found to increase with the strain. The shortest path is not static but changes dynamically with the strain. As shown in Figure 5, the path detours at 0.45 strain, and a new gap with shorter distance replaced the previous one.

To improve accuracy, a large data set is desired as the field intensification factor diverges near percolation. Therefore we repeated the simulation 90 times and obtained 180 sets of data in both the $y$ and $z$ direction. Given the potential for tunneling, we consider the system still percolated if the total distance between fillers is smaller than 2 nm. The probability of percolation is analyzed as a function of strain from the statistical data and a clear decreasing trend is observed as shown in Figure 6. Cautions are required in calculating the average field intensification factor, $L/d$, as it is important to consider the averaging effect in all three dimensions. For this purpose, the 180 sets of data are first divided into 60 groups to get an average value of $d/L$ within each group, and then the final $L/d$ is obtained by averaging the reciprocal of *d/L* of each group. This is equivalent to connecting 3 simulation boxes in series first and then connecting the stacked boxes in parallel, corresponding to a thin film geometry with an aspect ratio of 20:1. As shown in Figure 6, *L/d* exhibits a similar trend as the percolation probability and decreases dramatically with the increasing strain.

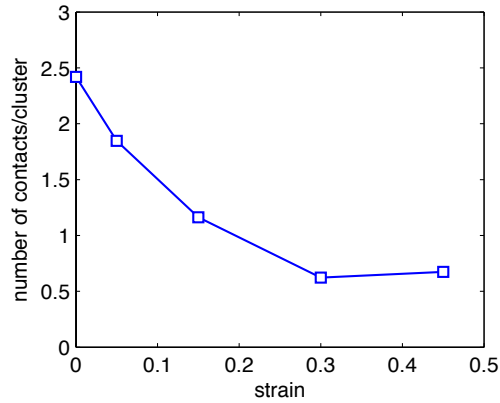

Figure 4 Number of contacts per cluster as a function of strain.

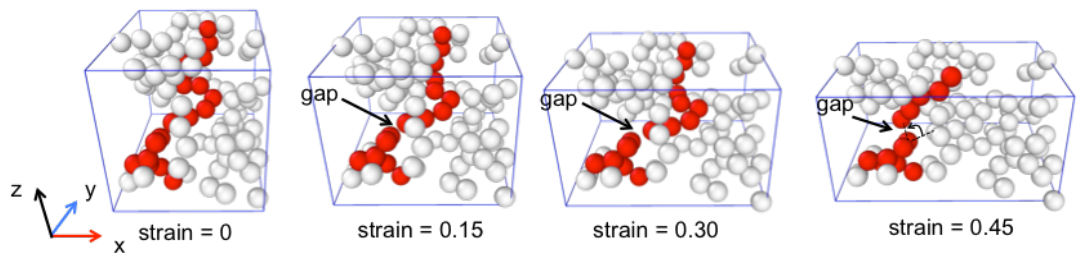

Figure 5 The simulated structure of the composites under different strain. The shortest path bridging the top and bottom surface is marked in red. The strain is along *x* direction and the field is along *y* or *z* direction.

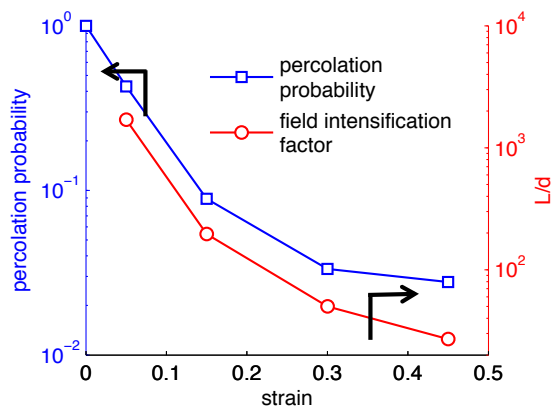

Figure 6 The calculated percolation probability and field intensification factor as a function of strain

The simulated dielectric spectra are plotted in Figure 7. The shape of the spectra resembles the one shown in Figure 1 with both real and imaginary permittivities decreasing with the strain. The relaxation intensity of MWS polarization, $\Delta\varepsilon$, is found to decrease with strain and the time constant also decreases. $\Delta\varepsilon$ can be analyzed from either the real or imaginary permittivity. For the real part, $\Delta\varepsilon$ is directly calculated as the difference between the high and low frequency permittivities, and the values are listed in Table 3. For a single time constant, the peak value of the imaginary permittivity is $\Delta\varepsilon/2$ as predicted by (4). However in this case, the loss peak is broadened due to the complex morphology of carbon black and the polarization time constant is a distribution. Nevertheless, $\Delta\varepsilon$ can still be obtained by analyzing the peak area, which is a constant with respect to $\Delta\varepsilon$,

$$\int_{\omega\tau\ll1}^{\omega\tau\gg1} \varepsilon'' d\log\omega = \int_{\omega\tau\ll1}^{\omega\tau\gg1} \sum_i \frac{\Delta\varepsilon_i \omega\tau_i}{1+\omega^2\tau_i^2} d\log\omega = \frac{\pi}{2}\sum_i \Delta\varepsilon_i = \frac{\pi}{2}\Delta\varepsilon \quad (8)$$

The results are also summarized in Table 3, which matches closely with the ones obtained from the real permittivity.

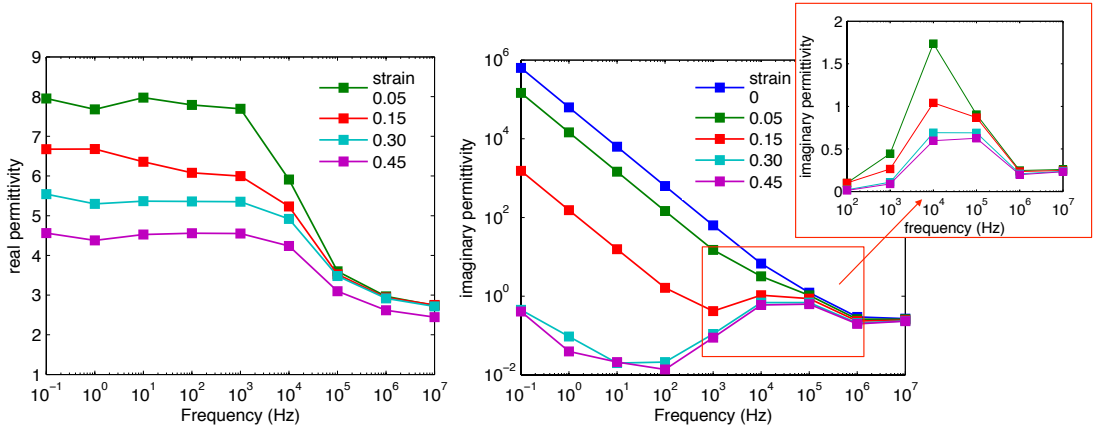

Figure 7 The simulated frequency spectra of the real and imaginary permittivity of the composites as a function of strain and frequency. The inset graph plots the MWS loss peak in a linear *y* axis, with dc conduction contribution subtracted, to better show the changes in the magnitude of MWS polarization.

Table 3 The relaxation intensity of MWS polarization calculated from the simulated real and imaginary permittivity respectively.

| strain | $\Delta\varepsilon$ from $\varepsilon'$ | $\Delta\varepsilon$ from $\varepsilon''$ |
|---|---|---|
| 0.05 | 5.2 | 5.0 |
| 0.15 | 3.9 | 3.7 |
| 0.30 | 2.8 | 2.5 |
| 0.45 | 2.1 | 2.2 |

The frequency and strain dependence of the real permittivity and loss tangent is plotted in comparison with the experimental results in Figure 8. The general trend is faithfully reproduced. The data in Figure 8 (a1) and (a2) does not show a loss peak as shown in Figure 1, possibly due to the shift of MWS polarization to higher frequencies at a lower volume fraction of 7.1 % with smaller dimension of the agglomerates[23]. Nevertheless, as shown in Figure 9, both simulated and measured imaginary permittivity shows a power-law declining behavior at 1 kHz with a rapid decrease at small strain and gradual leveling-off as strain increases further, consistent with change of structural parameters shown in Figure 6.

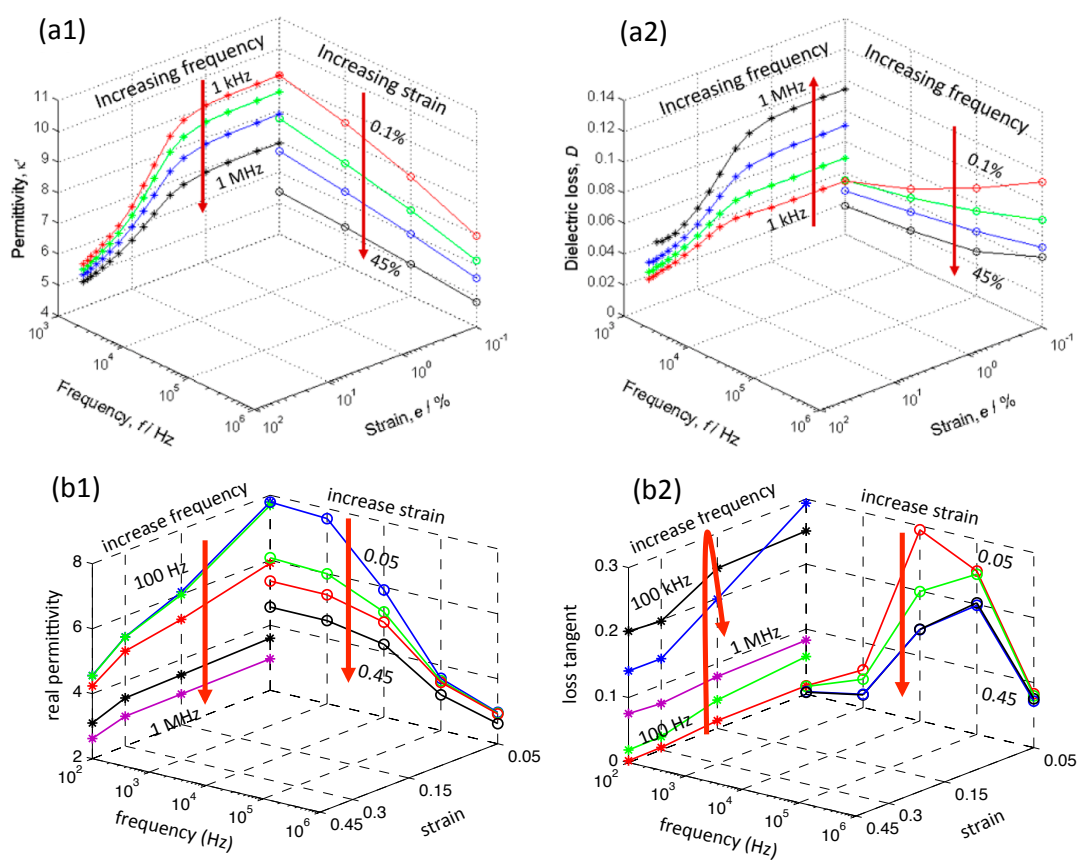

Figure 8 The strain and frequency dependence of real permittivity and loss tangent from experiment and simulation. (a1) (a2) are taken from Huang [*Macromolecules* **49**, 2339 (2016)], with a carbon black volume fraction of 7.1% (Figure is adapted with permission. Copyright ACS, 2016). (b1)(b2) are from the simulation results.

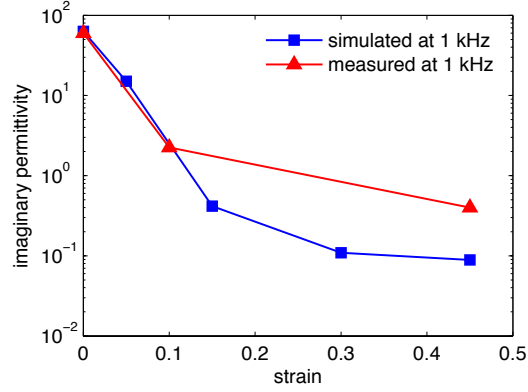

Figure 9 The imaginary permittivity of measured and simulated spectroscopy as a function of strain. The experimental data is taken from reference 1.

The discrepancy between the numerical values of simulated and measured spectra can be understood from the simulation setup. First, the simulation assumes an already percolated initial structure, which overestimates the dielectric loss as the percolation is a low probability event in real morphology. Second, the finite size of the simulation box limits the value of the field intensification factor, and thus underestimates the real permittivity. Here we would like to remind the readers that, the purpose of this paper is not to exactly reproduce the numerical data from the experiment, but rather to utilize the simulation to reveal the physical mechanism. The numerical accuracy, however, can be improved by tediously tuning the input material parameters and refining the simulation structure.

## 5. Conclusions

In this work, we demonstrate that the Maxwell-Wagner-Sillars (MWS) polarization is the major mechanism determining the dielectric permittivity of carbon black reinforced natural rubber in the measured frequency range. The large permittivity values are due to the long-range connectivity of the conductive filler network, which is able to enhance the electric field at the polymer by several orders of magnitude. Finite element simulations further reveals that stretching the polymer film can break the connection between filler clusters and thus alleviate the field intensification, which is responsible for the observed permittivity decrease. The virtually reproduced power-law permittivity-strain dependence is consistent with experimentally measured results. This investigation excludes any electronic interfacial effect as the necessary cause of the phenomenon, and on the contrary, suggests it rather as a bulk phenomenon. In fact, the similar hysteresis observed in the dielectric and mechanical responses or the "Payne" effect, indicates that they share the same physical origin—the dynamic breakage and recovery of the weak linkage between adjacent filler clusters[24]. From this point of view, the polymer-filler interactions will only have a secondary effect in this case by affecting the filler dispersion state or the load transfer between the filler and polymer. This conclusion can be extended to other elastomer composite systems containing conductive fillers.

**SUPPORTING INFORMATION**

The simulated stress and strain distribution in the composites. The simulated tensile stress of the composite as a function of strain.

## Acknowledgement

The authors would like to express their thanks to the financial support from the Office of Naval Research (N000141310173). The authors also thank Dr. Timothy Krentz at Savannah River National Laboratory, He Zhao, Yixing Wang and Dr. Catherine Brinson at Northwestern University, and Dr. Curt Breneman at Rensselaer Polytechnic Institute for their useful discussions.

## References

1. Huang, M.; Tunnicliffe, L. B.; Zhuang, J.; Ren, W.; Yan, H.; Busfield, J. J. C. Strain-Dependent Dielectric Behavior of Carbon Black Reinforced Natural Rubber. *Macromolecules* **2016,** 49, (6), 2339-2347.

2. Huang, X.; Zhi, C., *Polymer Nanocomposites: Electrical and Thermal Properties*. Springer International Publishing: Switzerland, 2016.

3. Koo, J. H., *Polymer Nanocomposites: Processing, Characterization, and Applications*. McGraw-Hill Professional Publishing: New York, 2006.

4. Calebrese, C.; Hui, L.; Schadler, L. S.; Nelson, J. K. A Review on the Importance of Nanocomposite Processing to Enhance Electrical Insulation. *IEEE Trans. Dielectr. Electr. Insul.* **2011,** 18, (4), 938-945.

5. Tanaka, T.; Kozako, M.; Fuse, N.; Ohki, Y. Proposal of a Multi-Core Model for Polymer Nanocomposite Dielectrics. *IEEE Trans. Dielectr. Electr. Insul.* **2005,** 12, (4), 669-681.

6.	Vaia, R. A.; Maguire, J. F. Polymer Nanocomposites with Prescribed Morphology: Going Beyond Nanoparticle-Filled Polymers. *Chem. Mater.* **2007,** 19, (11), 2736-2751.

7.	Fu, S.-Y.; Feng, X.-Q.; Lauke, B.; Mai, Y.-W. Effects of Particle Size, Particle/Matrix Interface Adhesion and Particle Loading on Mechanical Properties of Particulate–Polymer Composites. *Composites, Part B.* **2008,** 39, (6), 933-961.

8.	Ishii, H.; Sugiyama, K.; Ito, E.; Seki, K. Energy Level Alignment and Interfacial Electronic Structures at Organic/Metal and Organic/Organic Interfaces. *Adv. Mater.* **1999,** 11, (8), 605-625.

9.	Labardi, M.; Prevosto, D.; Nguyen, K. H.; Capaccioli, S.; Lucchesi, M.; Rolla, P. Local Dielectric Spectroscopy of Nanocomposite Materials Interfaces. *J. Vac. Sci. Technol., B: Nanotechnol. Microelectron.: Mater., Process., Meas., Phenom.* **2010,** 28, (3), C4D11-C4D17.

10.	Roy, M.; Nelson, J. K.; Schadler, L. S.; Zou, C.; Fothergill, J. C. In *The Influence of Physical and Chemical Linkage on the Properties of Nanocomposites*, , IEEE Conference on Electrical Insulation and Dielectric Phenomena (CEIDP) Nashville, USA, 16-19 Oct. 2005, pp 183-186.

11.	Nan, C.-W.; Shen, Y.; Ma, J. Physical Properties of Composites near Percolation. *Annu. Rev. Mater. Res.* **2010,** 40, 131-151.

12.	Huang, Y.; Krentz, T. M.; Nelson, J. K.; Schadler, L. S.; Li, Y.; Zhao, H.; Brinson, L. C.; Bell, M.; Benicewicz, B.; Wu, K.; Breneman, C. M. In *Prediction of Interface Dielectric Relaxations in Bimodal Brush Functionalized Epoxy Nanodielectrics by Finite Element Analysis Method*, IEEE Conference on Electrical Insulation and Dielectric Phenomena (CEIDP), Des Moines, USA, 19-22 Oct. 2014, pp 748-751.

13.	Hassinger, I.; Li, X.; Zhao, H.; Xu, H.; Huang, Y.; Prasad, A.; Schadler, L.; Chen, W.; Brinson, L. C. Toward the Development of a Quantitative Tool for Predicting Dispersion of Nanocomposites under Non-Equilibrium Processing Conditions. *J. Mater. Sci.* **2016,** 51, (9), 4238-4249.

14.	Zhao, H.; Li, X.; Zhang, Y.; Schadler, L. S.; Chen, W.; Brinson, L. C. Perspective: Nanomine: A Material Genome Approach for Polymer Nanocomposites Analysis and Design. *APL Mater.* **2016,** 4, (5), 053204.

15.	Meier, J. G.; Klüppel, M. Carbon Black Networking in Elastomers Monitored by Dynamic Mechanical and Dielectric Spectroscopy. *Macromol. Mater. Eng.* **2008,** 293, (1), 12-38.

16.	McLachlan, D.; Heaney, M. B. Complex Ac Conductivity of a Carbon Black Composite as a Function of Frequency, Composition, and Temperature. *Phys. Rev. B* **1999,** 60, (18), 12746.

17.	Bartnikas, R. In *Engineering Dielectrics*; American Society for Testing and Materials: Philadephia, 1983; Vol. IIA, p 61.

18. Klein, R. J.; Zhang, S.; Dou, S.; Jones, B. H.; Colby, R. H.; Runt, J. Modeling Electrode Polarization in Dielectric Spectroscopy: Ion Mobility and Mobile Ion Concentration of Single-Ion Polymer Electrolytes. *J. Chem. Phys.* **2006,** 124, (14), 144903.

19. Koga, T.; Takenaka, M.; Aizawa, K.; Nakamura, M.; Hashimoto, T. Structure Factors of Dispersible Units of Carbon Black Filler in Rubbers. *Langmuir* **2005,** 21, (24), 11409-11413.

20. Witten, T. A.; Sander, L. M. Diffusion-Limited Aggregation. *Phys. Rev. B* **1983,** 27, (9), 5686.

21. Boylan, J. Carbon - Graphite Materials. *Mater. World* **1996,** 4, (12), 707-8.

22. Monroe, D. Hopping in Exponential Band Tails. *Phys. Rev. Lett.* **1985,** 54, (2), 146.

23. Sillars, R. W., In *Electrical Insulating Materials and Their Application*. Peter Peregrinus Ltd. for the Institution of Electrical Engineers: Stevenage, 1973, p 56.

24. Wang, M.-J. The Role of Filler Networking in Dynamic Properties of Filled Rubber. *Rubber Chem. Technol.* **1999,** 72, (2), 430-448.